\documentclass{ws-ijmpd}
\usepackage[super,compress]{cite}
\begin{document}


%
\def\be{\begin{equation}}
\def\ee{\end{equation}}
\def\bea{\begin{eqnarray}}
\def\eea{\end{eqnarray}}
\def\p{\partial}
\def\n{\nabla}
\def\dkk{\p_k\p^k}
\def\sh{{\sigma}}
\def\Ps{{{\cal{P}}}}
\def\L{{\pounds}}
\def\R{{{\cal{R}}}}
\def\Va{{V_\alpha}}
\def\Q{{\cal{Q}}}
\def\J{{{\cal{J}}}}

\def\vp{{\varphi}}
\newcommand\dvp[1]{{\delta\varphi_{#1}}}
\newcommand\dvpI[1]{{\delta\varphi_{{#1}I}}}
\newcommand\dvpK[1]{{\delta\varphi_{{#1}K}}}
\newcommand\dvpM[1]{{\delta\varphi_{{#1}M}}}
\newcommand\dvpN[1]{{\delta\varphi_{{#1}N}}}
\newcommand\dvpL[1]{{\delta\varphi_{{#1}L}}}
\newcommand\dvpJ[1]{{\delta\varphi_{{#1}J}}}
\newcommand\dU[1]{{\delta U_{#1}}}
\def\S{{\cal{S}}}
\def\H{{\cal H}}
\def\cs2{c_{\rm{s}}^2}
\def\U0{{\bar U_0}}
\def\wt{\widetilde}
\def\dT{{\delta{\bf T}_1}}
\def\dTT{{\delta{\bf T}_2}}
\def\drho{{\delta\rho_1}}
\def\drhorho{{\delta\rho_2}}
\def\dP{{\delta P_1}}
\def\dPP{{\delta P_2}}
\def\dij{\delta_{ij}}
\def\12{\frac{1}{2}}
\def\BkBk{{B_{1,k}B_{1,}^{~k}}}
\def\ppij{{\p^{-1}_i\p^{-1}_j}}
\def\dvpdvpKll{\delta\vp_{1K,l}\delta\vp_{1K,}^{~~~~l}}
\def\Xkdvk{{\sum_K X_K \delta\vp_{1K}}}
\def\M{{\cal{M}}}
\def\k{{\bf{k}}}
\def\q{{\bf{q}}}
\def\kvi{{{k^i}}}
\def\qvi{{{q^i}}}
\def\pvi{{{p^i}}}
%


\def\tom{{\rm{tom}}}
\def\syn{{\rm{syn}}}
\def\com{{\rm{com}}}
\def\fg{{\rm{flat}}}
\def\lg{{\ell}}
\def\udg{{\delta\rho}}

\def\Mt{{\cal M}}
\def\Et{{\cal E}}

 \def\MI{{\cal M_{\rm{1}}}}
 \def\MII{{\cal M_{\rm{2}}}}
 \def\MIII{{\cal M_{\rm{3}}}}
 \def\EI{{\cal E_{\rm{1}}}}
 \def\EII{{\cal E_{\rm{2}}}}
 \def\EIII{{\cal E_{\rm{3}}}}

\markboth{Adam J.~Christopherson}
{Cosmological Perturbations: Vorticity, Isocurvature and Magnetic Fields}

\title{COSMOLOGICAL PERTURBATIONS: VORTICITY, ISOCURVATURE AND MAGNETIC FIELDS\footnote{Based on a seminar given at UC-Berkeley and various other US institutions in Autumn 2012.}}

\author{ADAM J.~CHRISTOPHERSON}

\address{School of Physics and Astronomy, University of Nottingham, University Park, Nottingham, NG7 2RD, UK\\
{achristopherson@gmail.com}}

\maketitle

\begin{history}
\received{Day Month Year}
\revised{Day Month Year}
\end{history}

\begin{abstract}
In this paper I review some recent, interlinked, work undertaken using cosmological perturbation theory -- a powerful technique for modelling inhomogeneities in the Universe. The common theme which underpins these pieces of work is the presence of non-adiabatic pressure, or entropy, perturbations. After a brief introduction covering the standard techniques of describing inhomogeneities in both Newtonian and relativistic cosmology, I discuss the generation of vorticity. As in classical fluid mechanics, vorticity is not present in linearized perturbation theory (unless included as an initial condition). Allowing for entropy perturbations, and working to second order in perturbation theory, I show that vorticity is generated, even in the absence of vector perturbations, by purely scalar perturbations, the source term being quadratic in the gradients of first order energy density and isocurvature, or non-adiabatic pressure perturbations. This generalizes Crocco's theorem to a cosmological setting. I then introduce isocurvature perturbations in different models, focusing on the entropy perturbation in standard, concordance cosmology, and in inflationary models involving two scalar fields. As the final topic, I investigate magnetic fields, which are a potential observational consequence of vorticity in the early universe. 
I briefly review some recent work on including magnetic fields in perturbation theory in a consistent way. I show, using solely analytical techniques, that magnetic fields can be generated by higher order perturbations, albeit too small to provide the entire primordial seed field, in agreement with some numerical studies. I close with a summary and some potential extensions of this work.
\end{abstract}



\tableofcontents

\section{Introduction}	

Cosmology, as a science, has greatly advanced over the last several decades. The first observational evidence that the Universe is expanding came from Hubble who noticed that galaxies were receding from us at a velocity proportional to their distance\cite{Hubble15031929}. Since then, there have been myriad observations made which have provided strong evidence supporting the Big Bang cosmological model. More recently, observations of distant supernovae showed that the linear relationship between distance and recession velocity is violated for the greatest distances, providing evidence that the Universe is accelerating, or at least appears to be, on the largest scales today\cite{Riess:1998cb, Perlmutter:1998np}. The cause for this acceleration is not known, and could be due to a yet undetected energy species, with negative pressure, dubbed dark energy\cite{Copeland:2006wr}, or could represent a modification to general relativity in low density environments\cite{Clifton:2011jh}.

Another piece of evidence strongly supporting the Big Bang theory is the  Cosmic Microwave Background (CMB) -- radiation formed early on when the Universe had expanded and cooled sufficiently to allow photons to stream --  first detected by Penzias and Wilson\cite{Penzias:1965wn}. Experiments since have used more and more advanced technology, enabling us to use observations of the CMB to strongly constrain models of the early universe. The Cosmic Background Explorer (COBE) was the first satellite to detect anisotropies in the CMB, and measured the radiation with a black body spectrum of $2.75 {\rm K}$, as predicted by theory\cite{COBE}. The spectrum of the anisotropies were then mapped out by the Wilkinson Microwave Anisotropy Probe (WMAP)\cite{WMAP7} and then, more recently, to a higher precision by Planck\cite{Ade:2013uln}.

 The data collected by these experiments have allowed us to rule out models of inflation  -- the period of accelerated expansion in the very early universe postulated to resolve problems of the original Hot Big Bang model. Based on their predictions for the spectral index of the primordial perturbations, the strength of the tensor perturbations as well as the non-Gaussianity of the initial fluctuations, the set of inflationary models can be constrained by observations. Very recently, the first measurement of the B-mode polarisation signal of the CMB was made by the Background Imaging of Cosmic Extragalactic Polarization (BICEP) experiment\cite{Ade:2014xna}. If the primordial origin of this signal is confirmed, it will be extremely strong evidence that the Universe went through a period of inflation. Further missions are planned to map the spectrum of B-mode polarisation, such as 
the Polarized Radiation Imaging and Spectroscopy Mission (PRISM\cite{Andre:2013nfa} ), the Experimental Probe of Inflationary Cosmology (EPIC\cite{Bock:2008ww}), LITEBird\cite{Matsumura:2013aja} and the Cosmic Origins Explorer (COrE\cite{Bouchet:2011ck}), among others. These will enable us to further tighten constraints on inflationary models.\\
 
In order to study the theoretical framework of the standard cosmological model, the Universe is well-described on the largest scales by the homogeneous and isotropic Friedmann-Lem\^itre-Robertson-Walker (FLRW) of general relativity. However, this cannot capture the complexity of all the structure in our Universe. To do this, we adopt a perturbative approach, introducing small, inhomogenous perturbations on top of the background FLRW solution. This is called \emph{cosmological perturbation theory}. Such an approach introduces gauge issues -- since general relativity is a covariant theory, choosing a particular background introduces spurious, unphysical degrees of freedom, called gauge artefacts. However, these can be removed, and are treated either on a case-by-case basis\cite{Lifshitz:1945du, Lifshitz:1963ps} or in a systematic manner\cite{Bardeen:1980kt}. There has been much theoretical progress on the latter technique resulting in linear \emph{metric} perturbation theory becoming a foundation of modern theoretical cosmology. Following on from Bardeen's seminal paper, two review articles were presented by Kodama and Sasaki\cite{ks} and Mukhanov, Feldman and Brandenberger\cite{mfb}. Arguably, these three articles together form the basis of linear metric cosmological perturbation theory.
\footnote{An alternative method -- the so-called covariant approach -- is also often studied which defines gauge invariant variables through the Stewart-Walker lemma\cite{Stewart:1974uz}. See, e.g., Refs.~\refcite{Ellis:1989ju},\refcite{Ellis:1989jt},\refcite{Ellis:1990gi} for details and Refs.~\refcite{Bruni:1992dg},\refcite{Malik:2012dr} for articles discussing the relationship between the metric and covariant approaches.} By expanding the perturbations order-by-order in a series, each order being smaller than the one before, the theory can be extended to higher order. Mathematical differences between linear and higher orders can result in qualitatively different observational signatures which, given the wealth of current and future data, can now help distinguish between cosmological models. Higher order perturbation theory has been studied by many authors in recent years\cite{Tomita, Bruni:1996im, Mukhanov1997, Acquaviva:2002ud, Malik2004, Noh:2004bc, Malik:2005cy, Nakamura:2010yg, Christopherson:2011hn}; see Refs.~\refcite{MW2008},\refcite{thesis} for detailed bibliographies.\\

In this article I will review some recent work on various complementary topics in cosmological perturbation theory including the generation of vorticity and magnetic fields at second order in cosmological perturbation theory, in addition to linear non-adiabatic pressure perturbations in the cosmic fluid and inflationary models. The common theme that underpins these pieces of work is the presence of non-adiabatic pressure perturbations. This work shows the importance and potential observational consequences on non-adiabatic pressure perturbations.

The paper is structured as follows: in the next section, we discuss how inhomogeneites are modelled in Newtonian and relativistic cosmologies and in Section~\ref{sec:vorticity}, we consider vorticity generation. In Section~\ref{sec:nonad}, we investigate non-adiabatic pressure perturbations in various physical systems and in Section~\ref{sec:mag} we review some recent work on magnetic field generation. In Section~\ref{sec:discuss} I will conclude, and give some possible extensions to this work.

\section{Modelling inhomogeneities}
\label{sec:modelling}

Although the Friedmann model is a good approximation on large scales,  our Universe is not exactly homogeneous and isotropic -- there exists structure, such as galaxies, stars and planets, and inhomogeneities in the CMB. In order to model these inhomogeneities we consider perturbations about a homogeneous `background' solution. In this case, we can express the energy density, for example, as
\be 
\rho(\vec{x},t)=\bar{\rho}(t)\Big(1+\delta(\vec{x},t)\Big)\,.
\ee
Here we denote background, homogeneous quantities with an overbar and have introduced the density contrast, $\delta(\vec{x},t)$, which is inhomogeneous (i.e. a function of time, $t$, and position, $\vec{x}$). We then invoke Newtonian mechanics, introducing a fluid velocity, $\vec{v}(\vec{x},t)$ and the Newtonian potential, $\Phi(\vec{x},t)$. The dynamics result in the fluid evolution equations\cite{peebles}
\begin{align}
&\dot{\delta}+\vec{\nabla}\cdot\Big[(1+\delta)\vec{v}\Big]=0\,,\\
&\dot{\vec{v}}+H\vec{v}+(\vec{v}\cdot\vec{\nabla})\vec{v}
=-\vec{\nabla}\Phi-\frac{\vec{\nabla} P}{\bar{\rho}(1+\delta)}\,,
\end{align}
where $H$ is the Hubble expansion parameter. Additionally, the Poisson equation relates the density contrast to the Newtonian potential as
\be 
\nabla^2\Phi=4\pi G \bar{\rho}a^2\delta\,.
\ee

The fluid evolution equations can be simplified by linearizing the system to give
\begin{align}
\label{eq:deltadot}
&\dot{\delta}+\vec{\nabla}\cdot\vec{v}=0\,,\\
\label{eq:vdot}
&\dot{\vec{v}}+H\vec{v}=-\vec{\nabla}\Phi-\frac{1}{\bar{\rho}}\vec{\nabla}\delta P\,,
\end{align}
where $\delta P$ is the perturbation to the pressure. One can gain more intuition about the physics by writing the system as a single, second order differential equation. On considering a barotropic fluid, where $P=P(\rho)$, the pressure perturbation can be related to the density perturbation through the introduction of the sound speed as
\be 
\delta P=c_{\rm s}^2\delta \rho\,.
\ee
Then, the system of Eqs.~(\ref{eq:deltadot}) and (\ref{eq:vdot}) reduces to the single, second order differential equation
\be
\label{eq:deltasecond}
\frac{\partial^2\delta}{\partial t^2}+2H\frac{\partial \delta}{\partial t}=4\pi G\bar{\rho}\delta+c_{\rm s}^2\nabla^2\delta\,.
\ee
From this equation, we can see that the density perturbations are affected by three physical processes. The second term on the left hand side of Eq.~(\ref{eq:deltasecond}) is the Hubble drag which causes a suppression of the perturbations due to the expansion of the Universe. The first term on the right hand side is the gravitational term which sources the growth of perturbations by the gravitational instability, and the second term on the right hand side is the pressure term.

Thus far, we have considered Newtonian physics, however dynamics of the Universe are governed by general relativity. In particular, we must use relativity to describe regions of high density, fluids moving at an appreciable fraction of the speed of light, or scales which are a substantial fraction of the horizon.

Einstein's field equations relate the geometry of the spacetime, encoded in the Einstein tensor, $G_{\mu\nu}$, to the matter content whose distribution is described by the energy-momentum tensor, $T_{\mu\nu}$:
\be 
G_{\mu\nu}=8\pi G T_{\mu\nu}\,.
\ee
So how do we proceed in incorporating inhomogeneities into a relativistic model? One could consider a fully inhomogeneous solution to the field equations, taking into account each inhomogeneity. This approach is incredibly difficult, in part because the majority of known solutions to Einstein's theory are relatively simple. Instead, we take heed from the Newtonian case, and expand around a homogeneous background solution. This results in cosmological perturbation theory, a successful and well-studied tool for modelling the inhomogeneities and anisotropies in our Universe. 

We consider perturbations to the energy density, in an analogous way to the Newtonian case, 
\be 
\rho(\vec{x}	,t)=\bar{\rho}(t)+\delta\rho(\vec{x},t)=\bar{\rho}(t)\Big(1+\delta(\vec{x},t)\Big)\,,
\ee
but now, through the Einstein equations, the perturbations in the matter content induce perturbations to the spacetime. This results in the spacetime metric being expanded as
\be 
g_{\mu\nu}(\vec{x},t)=g_{\mu\nu}^{(0)}(t)+\delta g_{\mu\nu}(\vec{x},t)\,.
\ee
Here, $g_{\mu\nu}^{(0)}(t)$ is the Friedmann-Lema\^itre-Robertson-Walker (FLRW) background solution and $\delta g_{\mu\nu}(\vec{x},t)$ is the inhomogeneous perturbation to the metric. These inhomogeneous perturbations can be expanded order-by-order in a series expansion. For example, the energy density perturbation is\footnote{For review articles on higher order perturbation theory see Refs.~\cite{MM2008, MW2008}.}
\be 
\delta \rho=\delta\rho_1+\frac{1}{2}\delta\rho_2+\cdots\,,
\ee
where we denote the order of perturbations with a subscript. The expansion demands that 
\be 
\delta\rho_1\ll\bar{\rho}\,, \,\,\,
\rm{and} \,\,\,\,
\delta\rho_2 < \delta\rho_1\,.
\ee
For the remainder of the article, we drop the subscripts where not necessary and when considering only linear perturbations.\\

The background solution has the line element
\be 
ds^2=-dt^2+a^2(t)\delta_{ij}dx^idx^j\,,
\ee
where $a(t)$ is the scale factor, and we have assumed flat spatial slices (so the spatial metric is just the kronecker delta), in agreement with observational evidence. The most general perturbations to the flat FLRW solution can be expressed as the line element
\be 
ds^2=-(1+2\phi)dt^2+2a(t)B_idx^idt+a^2(t)(\delta_{ij}+2C_{ij})dx^idx^j \,,
\ee
where the perturbed quantities can be decomposed into scalar, vector and tensor types, according to their transformation behavior on spatial hypersurfaces, as\cite{ks, mfb}
\be 
B_i=B_{,i}-S_i\,, \qquad C_{ij}=-\psi\delta_{ij}+E_{,ij}+F_{(i,j)}+\frac{1}{2}h_{ij}\,.
\ee
Here, $\phi$ is the lapse function, $\psi$ is the curvature perturbation and $B$ and $E$ make up the shear. $F_i$ and $S_i$ are the vector metric perturbations, and $h_{ij}$ is the transverse, traceless tensor, or gravitational wave, perturbation.

Unlike the Newtonian case, performing this split into a background and perturbed spacetime introduces a problem in relativistic perturbation theory, namely, the \emph{gauge problem}. While general relativity is covariant, this splitting is not a covariant process, and therefore doing so introduces coordinate artefacts, or spurious gauge modes. This problem can be solved, however, by working in terms of gauge invariant quantities -- that is, variables which do not change under a gauge transformation -- or, equivalently, by choosing to write the metric in a particular gauge. The systematic resolution was discovered by Bardeen\cite{Bardeen:1980kt}, and there have been many articles written on this topic in the past few decades (see Ref.~\refcite{thesis} for a comprehensive bibliography).

As an illustrative example, let us consider the quantity $\zeta$, the curvature perturbation on uniform density hypersurfaces defined as
\be 
\label{eq:defzeta}
-\zeta=\psi+H\frac{\delta\rho}{\dot{\bar{\rho}}}\,.
\ee
If we consider the transformation of the curvature perturbation and density perturbation under the coordinate transform
 \be 
 t\to\widetilde{t}=t-\delta t\,,
 \ee
 we obtain
\bea
\widetilde{\psi}&=&\psi-H\delta t \,,\\
\widetilde{\delta\rho}&=&\delta\rho+\dot{\rho}\delta t\,,
\eea
from which we can see immediately that $\zeta$, as defined in Eq.~(\ref{eq:defzeta}), is invariant under this transformation. In order to circumvent the gauge `problem', we can either work in a particular gauge or with a particular choice of gauge invariant variables; the two are equivalent.

There are several such gauges, or combinations of gauge invariant quantities, that one can choose to work with; an advantage of this `problem' is that one can choose a gauge most adaptable to the particular problem at hand. One useful choice is the so-called Newtonian gauge, in which the shear, $\sigma\equiv a(a\dot{E}-B)$, is zero, and the line element takes the form\footnote{Note that here we consider only scalar perturbations.}
\be 
ds^2=a^2(\eta)\Big[-(1+2\Phi)d\eta^2+(1-2\Psi)\delta_{ij}dx^idx^j\Big] \,.
\ee
The two gauge invariant scalar perturbations of the metric are now the Bardeen potentials, $\Phi$ and $\Psi$, and where we have introduced the conformal time, $\eta$, defined through $d\eta=dt/a$. In this gauge, the governing equations for the fluid come from energy-momentum conservation:
\begin{align}
&\delta'+(1+w)(\nabla^2v-3\Psi')=3\H(w-c_{\rm s}^2)\delta\,,\\
&v'+\H(1-3w)v+\frac{w'}{1+w}v+\frac{\delta P}{\bar{\rho}(1+w)}+\Phi=0\,.
\end{align}
Here, $v$ is the scalar potential of the fluid three-velocity, $w=\bar{P}/\bar{\rho}$ is the background equation of state, and 
$c_{\rm s}^2={\bar{P}}' / {\bar{\rho}}'$ is the adiabatic sound speed. We denote a derivative with respect to conformal time, $\eta$, with a dash. The Poisson equation, relating the potential to the matter variables is
\be 
\nabla^2\Psi=-4\pi Ga^2\bar{\rho}\Big[\delta-3\H(1+w)\nabla^2v\Big]\,,
\ee
which can be simplified to
\be 
\nabla^2\Psi=-4\pi Ga^2\delta_{\rm c}\,,
\ee
where we have defined the comoving density contrast, an alternative gauge-invariant variable,  as
\be 
\delta_{\rm c}\equiv\delta-3\H(1+w)\nabla^2v\,.
\ee
In the case of pressureless dust, for which $w=0=c_{\rm s}^2$, the above set of equations reduces to the simple second order differential equation\cite{Christopherson:2012kw}
\be
\delta_{\rm c}''+\H\delta_{\rm c}'=4\pi Ga^2\bar{\rho}\delta_{\rm c}\,.
\ee
For non-relativistic matter, this is equivalent to Eq.~(\ref{eq:deltasecond}), thus explaining the origin for the name Newtonian gauge. 
Another choice of gauge, particular useful for calculations of perturbations during inflation, and one which we will use later in this article, is the uniform curvature, or flat gauge. This gauge choice consists of unperturbed spatial slices, such that $E=\psi=0$, and the line element takes the form
\be 
ds^2=a^2(\eta)\Big[-(1+2\phi)d\eta^2+B_{,i}dx^id\eta+\delta_{ij}dx^i dx^j\Big] \,,
\ee
with the two gauge invariant variables in this gauge, $\phi$ and $B$. \\

In this section, I have introduced methods for modeling inhomogeneities in both Newtonian and relativistic cosmologies. While the former is more simple, and describes some regimes well, it is not sufficient for many purposes in cosmology. In particular, in order to model perturbations to relativistic species, such as neutrinos or photons, we must invoke general relativity, as is the case when considering regions of high pressure, such as the early universe, or when considering systems with dynamical dark energy\cite{Christopherson2010}. Additionally, when we are considering large regions, of a size comparable to the horizon size, relativistic effects are likely to become important. Therefore, in order to model the early universe, including the inflationary era and the formation of the cosmic microwave background, we must use relativistic perturbation theory. Similarly, although structure formation takes place in the Newtonian regime, there could be sizeable relativistic effects to the initial condition generation. Determining how to interpret Newtonian results from a relativistic point of view is an important problem which is starting to be addressed\cite{Green:2011wc, Chisari:2011iq, Haugg:2012ng, Bruni:2013qta, Hidalgo:2013mba}.

\subsection{Governing equations in the flat gauge}

In this section I will present the governing equations obtained from the energy-momentum conservation and Einstein equations in the uniform curvature gauge, which will be used to perform the calculations for the remainder of the article. From conservation of energy-momentum, we obtain an energy conservation equation
\be 
\delta\rho'+3\H(\delta\rho+\delta P)+(\bar{\rho}+\bar{P})\nabla^2v=0\,,
\ee
and from the momentum conservation equation
\be 
V'+\H(1-3c_{\rm s}^2)V+\frac{\delta P}{\bar{\rho}+\bar{P}}+\phi=0\,,
\ee
where we have introduced the velocity perturbation $V=v+B$. From the Einstein equations, we obtain
\begin{align}
&3\H^2\phi+\H\nabla^2B=-4\pi Ga^2\delta\rho\,,\\
&\H\phi=-4\pi Ga^2(\bar{\rho}+\bar{P})V\,,\\
&\H\phi'+\phi\Big(\frac{2a''}{a}-\H^2\Big)=4\pi Ga^2\delta P\,,\\
&B'+2\H B+\phi=0\,.
\end{align}
In order to write this in a form for easy solution later, we define
\be 
\mathcal{V}\equiv(\bar{\rho}+\bar{P})V\,,
\ee
for which the energy conservation equations Eqs.~(\ref{eq:encons}) and (\ref{eq:momcons}) can be reduced, using the field equations and again introducing the equation of state parameter, $w$, defined above, to
\begin{align}
\label{eq:encons}
&\delta\rho'+\frac{3}{2}\H(3+w)\delta\rho+3\H\delta P-k^2\mathcal{V}-\frac{9}{2}\H^2(1+w)\mathcal{V}=0\,,\\
\label{eq:momcons}
&\mathcal{V}'+\frac{\H}{2}(5-3w)\mathcal{V}+\delta P=0\,,
\end{align}
where we have switched to Fourier space, denoting the wavenumber by $k$, such that $\nabla^2=-k^2$. These equations will be solved, under certain approximations, in the later sections.

\section{Vorticity}
\label{sec:vorticity}

Vorticity is a phenomenon which is very common in engineering and applications of physics, and is extremely well studied in fluid dynamics\cite{landau}. Vorticity has recently been investigated in astrophysics situations\cite{DelSordo:2010mt}, however the role of vorticity in the early universe and cosmology has not been studied in detail. Since the early universe is a very complicated phase, one might expect that it is highly turbulent, and therefore that vorticity will play an important role. In this section, we investigate vorticity in the early universe.

In classical fluid dynamics, vorticity is defined as the curl of the fluid  velocity vector, which can be thought of as the circulation of the fluid per unit area at a point in the flow: 
\be 
\label{eq:vordef}
\vec{\omega}\equiv\vec{\nabla}\times\vec{v}\,.
\ee
Now, considering an inviscid fluid in the absence of gravity and other body forces and in a static universe, the evolution of the fluid velocity in Newtonian mechanics is governed by the Euler equation
\be 
\label{eq:euler}
\dot{\vec{v}}+(\vec{v}\cdot\vec{\nabla})\vec{v}=-\frac{1}{\rho}\vec{\nabla}P\,,
\ee
where $\rho$ is the energy density and $P$ the pressure of the fluid. Using Eq.~(\ref{eq:euler}), and the definition of the vorticity vector, Eq.~(\ref{eq:vordef}), the vorticity can be shown to evolve according to 
\be 
\dot{\vec{\omega}}=\vec{\nabla}\times(\vec{v}\times\vec{\omega})
+\frac{1}{\rho^2}\vec{\nabla}\rho\times\vec{\nabla}P\,.
\ee
The source term of this equation -- the second term on the right hand side -- is the baroclinic term. This is zero if gradients of the pressure and energy density are parallel, or if either the pressure or energy density are constant. This is the case for a barotropic fluid, for which the equation of state is of the form $P=P(\rho)$. Thus, we see that vorticity can only be sourced if we allow for a more general equation of state, depending on two independent variables, such as $P=P(\rho,S)$. This was first discovered by Crocco\cite{crocco}, who showed that allowing for entropy, $S$, provides a source for the vorticity.

Using this more general equation of state for the fluid, we can expand the pressure perturbation as\cite{nonad}
\be 
\delta P = \frac{\partial P}{\partial S}\delta S+\frac{\partial P}{\partial\rho}\delta\rho\,,
\ee
which can also be expressed as 
\be 
\label{eq:defdPnad}
\delta P=\delta P_{\rm nad}+c_{\rm s}^2\delta\rho\,,
\ee
on introducing the non-adiabatic pressure perturbation, $\delta P_{\rm nad}$, and the square of the adiabatic sound speed $c_{\rm s}^2$. This can be extended to higher order in cosmological perturbation theory\cite{Malik2004}. We will see that $\delta P_{\rm nad}$ is an important quantity for vorticity generation. We will study non-adiabatic pressure perturbations in more detail and in specific models in Section~\ref{sec:nonad}.

\subsection{Vorticity in cosmology}

In general relativity, vorticity is described by the projected, anti-symmetric covariant derivative of the fluid four-velocity:
\be 
\omega_{\mu\nu}=\frac{1}{2}p_\mu{}^\alpha p_\nu{}^\beta\Big(\nabla_\beta u_\alpha-\nabla_\alpha u_\beta\Big)\,,
\ee
where $p_{\mu\nu}$ projects into the fluid rest frame:
\be 
p_{\mu\nu}=g_{\mu\nu}+u_\mu u_\nu\,.
\ee
A vorticity vector can be constructed, by invoking the totally symmetric permutation tensor in the fluid rest frame, $\epsilon_{\mu\nu\delta}\equiv u^\gamma\epsilon_{\gamma\mu\nu\delta}$, to give\cite{Lu2008, Christopherson:2010dw}
\be 
\omega_\mu=\frac{1}{2}\epsilon_{\mu\nu\gamma}\omega^{\nu\gamma}\,.
\ee
Working now with only scalars and vector perturbations, the fluid four velocity is
\be 
u_\mu=-\frac{1}{2}a\Big[2(1+\phi_1)+\phi_2-\phi_1^2+v_{1k}v_1^k,-2V_{1i}-V_{2i}+2\phi_1B_{1i}\Big]\,,
\ee
from which we can obtain the components of the vorticity tensor, at first order
\be 
\omega_{1ij}=aV_{1[i,j]}\,,
\ee
and at second order
\be 
\omega_{2ij}=aV_{2[i,j]}+2a\Big[V_{1[i}'V_{1j]}+\phi_{1,[i}(V_1+B_1)_{j]}-\phi_1B_{1[i,j]}\Big]\,,
\ee
where square brackets around subscript indices denote antisymmetrization.

An evolution equation for the vorticity is obtained by taking the time derivative of the components of the tensor, order by order, and using the governing equations presented above to simplify. Doing so yields the simple result at first order,
\be 
\omega_{1ij}'-3\H c_{\rm s}^2\omega_{1ij}=0\,,
\ee
which is the well-known result\cite{ks} that, in the absence of anisotropic stress, the vorticity decays like $a^{-2}$ in a radiation dominated universe. In particular, if vorticity is initially zero, then it cannot be sourced in linear perturbation theory.

 At second order, it is trickier to obtain a closed equation; one must use the linear constraint and evolution equations, in addition to the second order momentum conservation equation. Omitting the details (see Refs.~\refcite{thesis},\refcite{vorticity}  for full details of the calculation), we show that vorticity is sourced even when neglecting linear vorticity,
\be 
\label{eq:vortev}
\omega_{2ij}'-3\H c_{\rm s}^2\omega_{2ij}
=\frac{2a}{\bar{\rho}+\bar{P}}\Bigg(3\H V_{1[i}\delta P_{\rm nad 1,j]}+\frac{1}{\bar{\rho}+\bar{P}}\delta\rho_{1[,j}\delta P_{\rm nad 1,i]}\Bigg)\,.
\ee
In analogy with classical fluid dynamics, the vorticity is sourced by a coupling between gradients of energy density and entropy perturbations. Similarly, for the case of a barotropic fluid, there is no vorticity generation even at second order, in agreement with Ref.~\refcite{Lu2008}.

Thus, we can see that the crucial input to determining the size of the effect of vorticity is the entropy perturbation. Unfortunately, it is nontrivial to compute the entropy perturbations, since they are dependent on the model of the early universe, encoded in the equation of state of the system. As an initial estimate, computed in Ref.~\refcite{Christopherson:2010ek}, we can take power law input spectra and perform an analytical calculation.\footnote{We note that, in Ref.~\refcite{Christopherson:2010ek}, incorrect input spectra was used for the density perturbation -- I have corrected this in the following, and therefore present an updated estimate.} In order to do so, we approximate Eq.~(\ref{eq:vortev}) by neglecting the first source term. We can then solve the system of Eqs.~(\ref{eq:encons}), (\ref{eq:momcons}), by seeking a power law solution for the energy density to give
\be 
\delta\rho_1(k,\eta)=A\Big(\frac{\eta}{\eta_0}\Big)^{-4}\,,
\ee
and then estimate the non-adiabatic pressure input power spectrum as
\be 
\delta P_{\rm nad 1}(k,\eta)=D\Big(\frac{k}{k_0}\Big)^2\Big(\frac{\eta}{\eta_0}\Big)^{-5}\,,
\ee
where the amplitudes are then set to CMB observations. Using these inputs, the evolution equation for the vorticity can be solved, to give a simple spectrum
\begin{align}
P_\omega(k,\eta)=\eta^2\ln^2\Big(\frac{\eta}{\eta_0}\Big)
&\Bigg[6.546\times 10^{-20}
\Bigg(\frac{k_c}{{\rm Mpc}^{-1}}\Bigg)^{12}\Big(\frac{k}{k_0}\Big)^9\Bigg]{\rm Mpc}^2\,,
\end{align}
where we have introduced a UV-cutoff $k_c$ -- this is required since nonlinear astrophysical processes will dominate the cosmological calculation on small scales, due to the breakdown of perturbation theory. We can see that the spectrum is blue
(with more power on smaller scales), but depends heavily on the small scale cutoff $k_c$. On substituting a realistic value for the cutoff, $k_c=10{\rm Mpc}^{-1}$, we obtain
\begin{align}
P_\omega(k,\eta)=\eta^2\ln^2\Big(\frac{\eta}{\eta_0}\Big)
&\Bigg[6.546\times10^{-8}\Big(\frac{k}{k_0}\Big)^9\Bigg]{\rm Mpc}^2\,.
\end{align}
Although this is a simple toy model, the spectrum has a non-negligible amplitude. This indicates that vorticity may be important on some cosmological scales, thereby motivating its further study.

Vorticity can have several observational effects; potentially the most interesting effect may come from the polarization of the CMB. The CMB radiation has two polarizations characterised, in analogy with electromagnetism, as B- and E-mode -- the B-mode is divergence free and the E-mode is curl free. In linear perturbation theory, scalar perturbations only generate E-mode polarization and although vector perturbations can generate B-mode polarization, they decay with the expansion of the Universe. Therefore, B-mode polarization is only produced, in the standard cosmological model, in the linear theory by tensors, or gravitational waves. However, second order vector perturbations, like vorticity formed by first order density and entropy perturbations, can source B-mode polarization. Therefore, it is expected that vorticity is important for current and future CMB polarization experiments.\footnote{Since this work was presented, an article\cite{Fidler:2014oda} has appeared in the literature suggesting that the B-mode signal from vorticity and other intrinsic effects is not substantial enough to contaminate future gravitational wave searches.}

\section{Entropy perturbations}
\label{sec:nonad}

We have seen in the previous section the importance of the non-adiabatic pressure perturbation\footnote{We will use the terms entropy, non-adiabatic pressure and isocurvature perturbations interchangeably throughout this article.} in sourcing vorticity. We will now go on to discuss entropy perturbations in more detail.

Non-adiabatic pressure perturbations are present in physical systems with an equation of state that depends upon two independent variables -- i.e. $P=P(\rho, S)$. So, a single barotropic fluid cannot support entropy perturbations. Similarly, a single scalar field (in the superhorizon limit) does not generate non-adiabatic pressure perturbations, as it can be treated as a barotropic fluid\cite{nonad, Arroja:2010wy}. This means that, in order to generate non-adiabatic pressure perturbations, we require either a single fluid with an equation of state more general than that of a barotropic fluid, or a system containing a collection of barotropic fluids, or scalar fields. Additionally, since non-adiabatic pressure perturbations are invariant under gauge transformations, they cannot be 'gauged-away'; that is, a gauge choice cannot be made such that they are zero. This highlights the physical importance of entropy perturbations.

The non-adiabatic pressure perturbation defined above in Eq.~(\ref{eq:defdPnad}) can be further decomposed into a part intrinsic to each fluid ($\delta P_{\rm intr}$) and a relative part between each fluid ($\delta P_{\rm rel}$). For a system of barotropic fluids the former is zero. The latter is due to the relative entropy perturbation between different fluids defined as
\be 
S_{\alpha\beta}=-3\H\Bigg(\frac{\delta\rho_\alpha}{\bar{\rho}_{\alpha}'}-\frac{\delta\rho_\beta}{\bar{\rho}_{\beta}'}\Bigg)\,,
\ee
where the relative contribution to the non-adiabatic pressure is then
\be 
\delta P_{\rm rel}=-\frac{1}{6\H \bar{\rho}'}\sum_{\alpha,\beta}
\bar{\rho}_{\alpha}'\bar{\rho}_{\beta}'
(c_\alpha^2-c_\beta^2)S_{\alpha\beta}\,.
\ee
Putting the two together, then results in
\be 
\delta P_{\rm rel}=\frac{1}{2\bar{\rho}'}\sum_{\alpha,\beta}(c_\alpha^2-c_\beta^2)(\bar{\rho}_{\beta}'\delta\rho_\alpha-\bar{\rho}_{\alpha}'\delta\rho_\beta)\,.
\ee

We will now study entropy perturbations in two different settings. First, we investigate the relative entropy perturbation in the usual cosmic fluid (i.e. baryons, cold dark matter, radiation, neutrinos...) between the relativistic and non-relativistic species. Then we will go on to study isocurvature perturbations in multiple field inflationary models, focusing on two different potentials which give qualitatively different results.

\subsection{Non-adiabatic pressure in the cosmic fluid}

In the standard cosmological model, anisotropies are sourced by quantum fluctuations from the inflationary epoch. The dynamics of inflation then determine the initial conditions of the primordial perturbations, either adiabatic, isocurvature or a mixture of the two. After the initial conditions have been imposed, there is no constraint on the contribution of the non-adiabatic pressure through observations of, e.g., the power spectrum.

The concordance cosmological model is a multi-component system consisting of several barotropic fluids. In Ref.~\refcite{Brown:2011dn}, we numerically compute the contribution to the non-adiabatic pressure perturbation arising from the relative entropy perturbation in the cosmic fluid. There are five fluids making up the concordance cosmology: radiation, or photons ($\gamma$), baryonic matter ($b$), cold dark matter (CDM, $c$), neutrinos ($\nu$) and dark energy, which we assume in this work to be a cosmological constant. The baryons and cold dark matter are pressureless, $w_b=w_c=c_b^2=c_c^2=0$, and the photons and neutrinos are relativistic, $w_\gamma=w_\nu=c_\gamma^2=c_\nu^2=\frac{1}{3}$.
The governing equations are presented in the synchronous gauge\cite{Ma:1995ey} and the extra gauge freedom is fixed by choosing to be comoving with the cold dark matter. 

\begin{figure}
\begin{center}
\includegraphics[width=0.6\columnwidth]{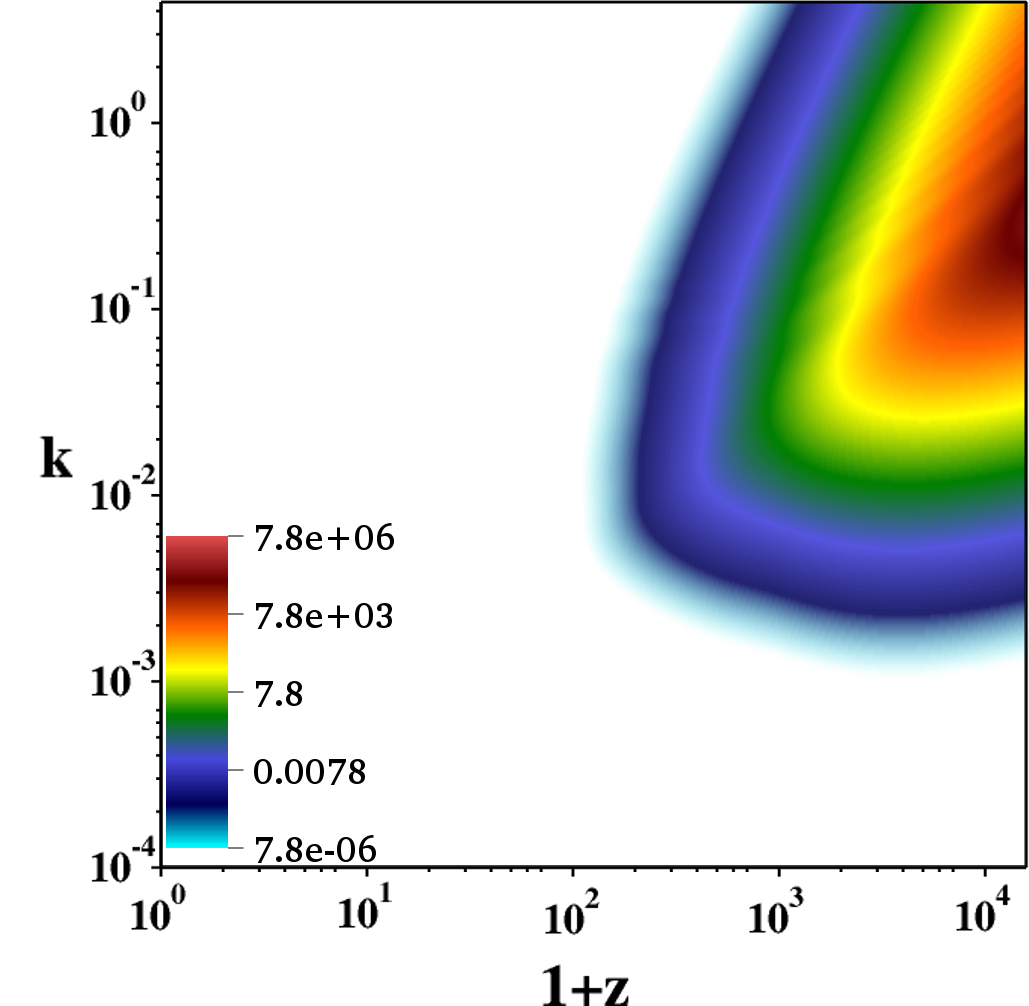}\end{center}
\caption{The power spectrum of the non-adiabatic pressure perturbation plotted against scale and redshift, reproduced with permission from Brown et al (2012), their figure 2.}
\label{fig:DeltaPRel}
\end{figure}

\begin{figure}
\begin{center}\includegraphics[width=0.6\columnwidth]{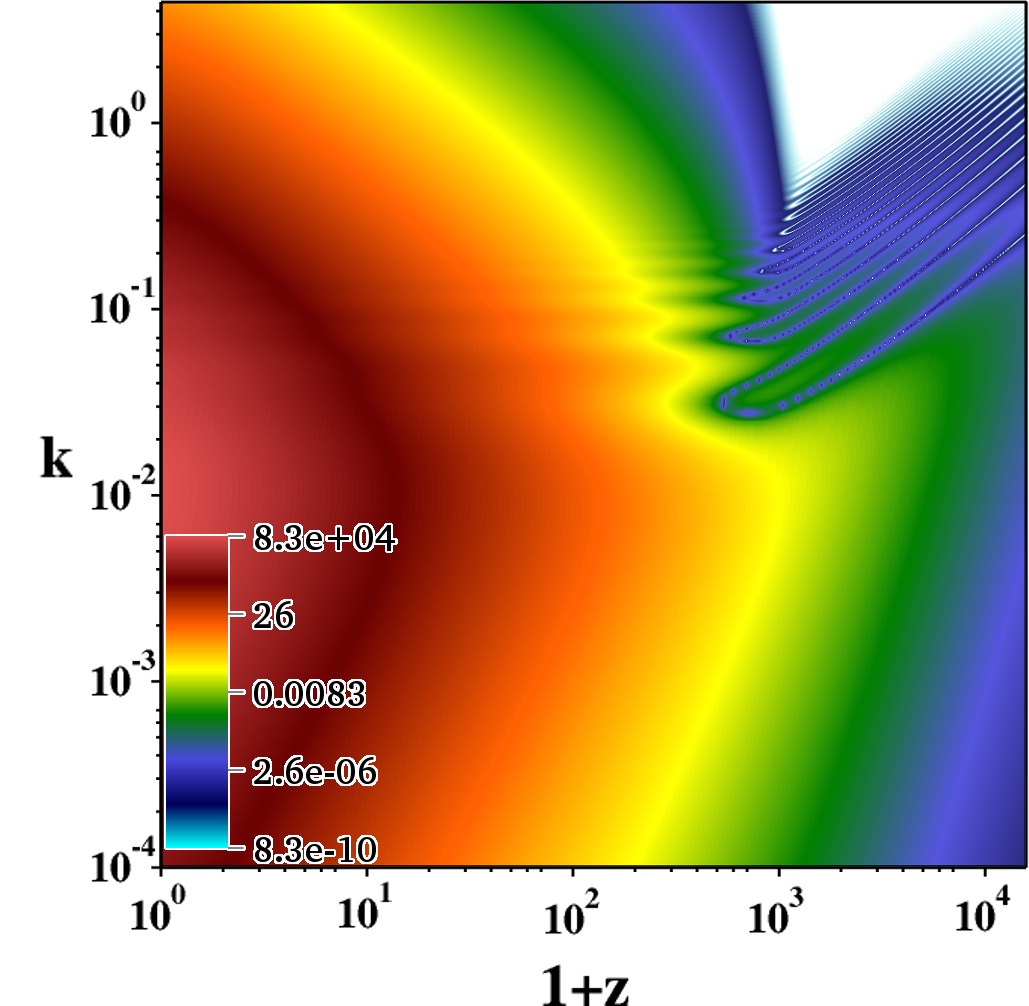}\end{center}
\caption{The power spectrum of the baryon density contrast, reproduced with permission from Brown et al (2012), their figure 1.}
\label{fig:BaryonPert}
\end{figure}

The initial conditions are taken to be adiabatic, such that the initial entropy perturbation is zero:
\begin{align}
&\delta_\gamma=\delta_\nu=\frac{4}{3}\delta_b=\frac{4}{3}\delta_c=-\frac{2}{3}Ck^2\eta_i^2\,,\\
&v_\gamma=v_b=-\frac{1}{18}Ck^2\eta_i^3\,, \\
&v_\nu=\frac{23+4R_\nu}{15+4R_\nu}v_\gamma\,, \,\,\, v_c=0\,,
\end{align}
where $R_\nu$ is the relative neutrino abundance.
We then use a modified version of CMBFast\cite{Seljak:1996is} to compute the relative non-adiabatic pressure perturbation. The  equations governing the system are presented in detail in Ref.~\refcite{Brown:2011dn}. In Fig.~\ref{fig:DeltaPRel}, we plot the power spectrum of the non-adiabatic pressure perturbation in terms of wavenumber, $k$, and redshift, $z$. 
We can see that $\delta P_{\rm rel}$ peaks around matter-radiation equality, and is orders of magnitude larger on small scales than on larger scales at that time. After matter-radiation equality, the relative entropy perturbation decays rapidly across all wavenumbers. In Fig.~\ref{fig:BaryonPert}, we plot the power spectrum of the baryon density contrast, for comparison.

\begin{figure}
\begin{center}
\includegraphics[width=\columnwidth]{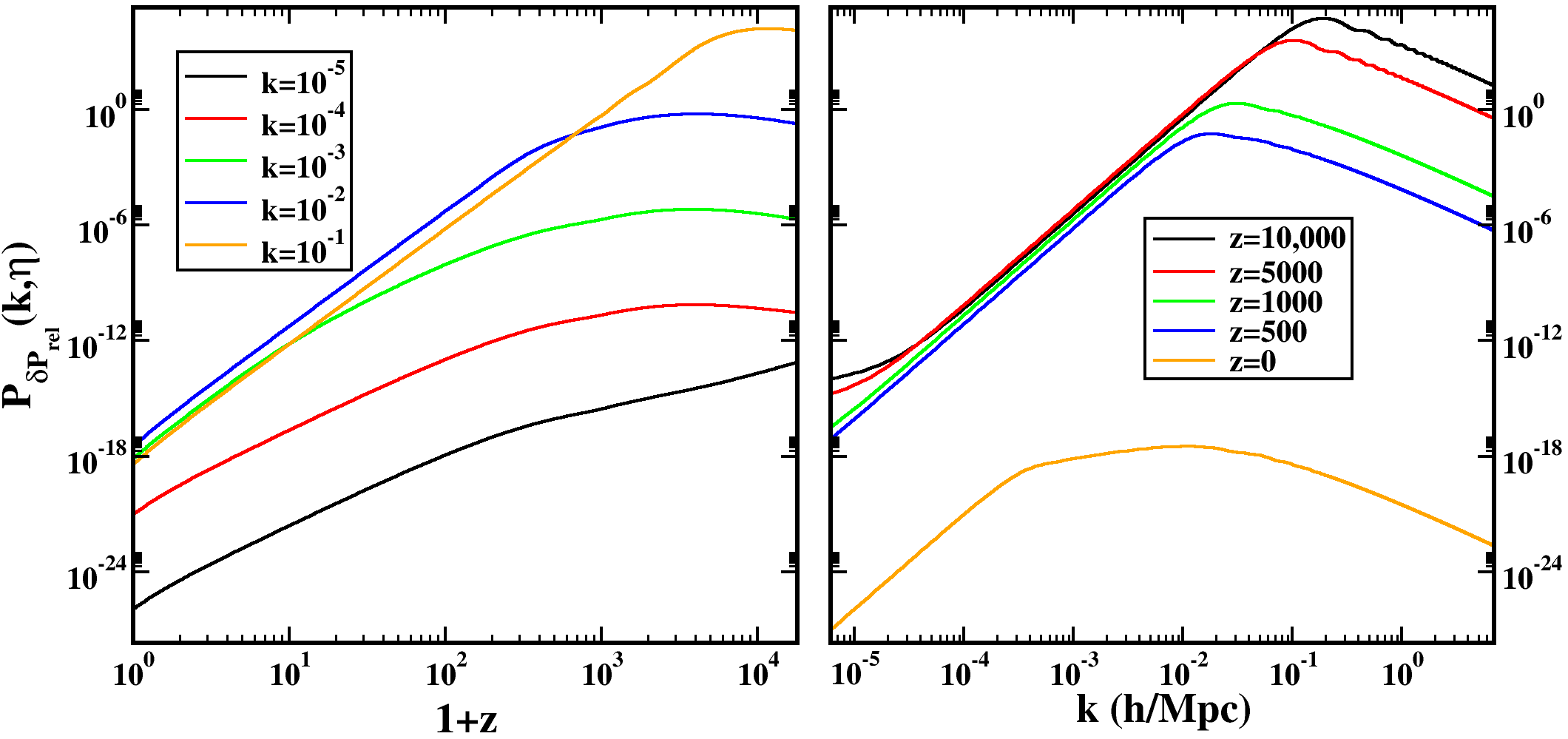}\end{center}
\caption{The power spectrum of the non-adiabatic pressure perturbation as a function of redshift for set wavenumber (left) and as a function of wavenumber for set redshift (right), reproduced with permission from Brown et al (2012), their figure 3.}
\label{fig:DeltaPRelKZ}
\end{figure}

In Fig.~\ref{fig:DeltaPRelKZ}, we plot the power spectrum of the non-adiabatic pressure as a function of redshift for a fixed wavenumber and as a function of wavenumber for a fixed redshift. Looking at the left panel, we can see the growth of the non-adiabatic pressure at early times, and the decay at later times. The peak is close to matter-radiation equality, with shorter wavelengths peaking earlier than longer wavelengths and decaying faster (for short wavelengths, $\delta P_{\rm rel}\propto a^{-3}$ while $\delta P_{\rm rel}\propto a^{-5/2}$ for long wavelengths). In the right hand panel, we can clearly see the decay of the non-adiabatic pressure towards to current epoch. Additionally, the mode with most power shifts to larger scales as we move toward the present day.\\

Here we have calculated the non-adiabatic pressure perturbation arising from the relative entropy between the different fluids in the concordance cosmology. We find a non-vanishing signature, even when the initial conditions were adiabatic, which grows at early times and peaks around matter-radiation equality. This entropy perturbation is already present in CMB calculations, yet has not been explicitly calculated before Ref.~\refcite{Brown:2011dn}. Although this does not influence the linear Boltzmann codes, the perturbation will have various effects. In particular, in Section~\ref{sec:vorticity} we saw how non-adiabatic pressure perturbations can source vorticity at second order in perturbation theory. This will be studied in a future publication.

\subsection{Isocurvature perturbations in multi-field inflationary models}

Inflation was initially postulated to solve the problems of the big bang cosmology, the horizon and flatness problems \cite{LLBook}. However, in addition to this, inflation also provides a mechanism for generating the perturbations which give rise to the anisotropies and inhomogeneities in the CMB. Quantum fluctuations in the inflationary field are amplified by the accelerated expansion of the universe, and once pushed outside the horizon become the classical seeds of structure formation.

While the inflationary paradigm is now well accepted, there is no one model that stands out as preferable. Although CMB observations are compatible with adiabatic initial conditions, such as those coming from inflation driven by a single scalar field, there is still some room for isocurvature perturbations which arise from inflationary models with multiple scalar fields. While the primordial power spectrum is predominantly adiabatic, there is still some room for a non-zero, albeit subdominant, isocurvature contribution\cite{WMAP7, Kawasaki:2007mb, Beltran:2008aa, Li:2010yb, Ade:2013uln, Savelainen:2013iwa}. {\sc Wmap} defines the parameter $\alpha(k_0)$ as
\be 
\label{eq:alphadef}
\frac{{\mathcal{P}}_{\mathcal S}}{{\mathcal{P}}_{\mathcal R}}=\frac{\alpha(k_0)}{1-\alpha(k_0)}\,,
\ee
and then the current data constrain $\alpha(k_0)\sim 0.13$. In this section, I will present work from Ref.~\refcite{Huston:2011fr} in which we consider non-adiabatic pressure perturbations from different multi-field inflationary models. 

We will use cosmological perturbation theory throughout, instead of the often-used $\delta N$ formalism \cite{Sasaki:1995aw, Lyth:2001nq, Lyth:2004gb}. By using perturbation theory, we will be able to consider isocurvature perturbations during inflation, and obtain the spectrum of the non-adiabatic pressure present at the end of inflation.

The two-field inflationary models that we will consider have the Lagrangian
\be 
{\mathcal{L}}=\frac{1}{2}\Big(\dot{\varphi}^2+\dot{\chi}^2\Big)
+V(\varphi,\chi)\,,
\ee
with two different choices for the potential $V(\varphi,\chi)$, and will compute perturbations working in the uniform curvature gauge. The energy density and pressure perturbations can be expressed in terms of the fields as\cite{Hwang:2001fb, Huston:2011fr}
\bea
\delta\rho&=&\sum_\alpha\Big(\dot{\varphi}_\alpha\dot{\delta\varphi}_\alpha
-\dot{\varphi}_\alpha^2\phi+V_{,\alpha}\delta\varphi_\alpha\Big)\,,\\
\delta P&=&\sum_\alpha\Big(\dot{\varphi}_\alpha\dot{\delta\varphi}_\alpha
-\dot{\varphi}_\alpha^2\phi-V_{,\alpha}\delta\varphi_\alpha\Big)\,,
\eea
and so we can compute the non-adiabatic pressure perturbation defined above in Eq.~(\ref{eq:defdPnad}). In order to compare the results to the comoving curvature perturbation, ${\mathcal{R}}$ which, for these models, takes the form
\be 
{\mathcal{R}}=\frac{H}{\dot{\varphi}^2+\dot{\chi}^2}\Big(\dot{\varphi}\delta\varphi
+\dot{\chi}\delta\chi\Big)\,,
\ee
 we use the comoving entropy perturbation defined as
\be 
{\mathcal{S}}=\frac{H}{\dot{\bar{P}}}\delta P_{\rm nad}\,.
\ee
An alternative, popular way of computing perturbations in two field inflation models is to rotate the fields into and adiabatic and isocurvature field, as was first performed in Ref.~\refcite{Gordon:2000hv}:
\bea
\delta\sigma&=&\cos\theta\delta\varphi+\sin\theta\delta\chi\,,\\
\delta s&=&-\sin\theta\delta\varphi+\cos\theta\delta\chi\,.
\eea
A comoving isocurvature perturbation can then be defined in terms of these variables as
\be 
\widetilde{\mathcal{S}}=\frac{H}{\dot{\sigma}}\delta s\,.
\ee
${\mathcal{S}}$ and $\widetilde{\mathcal{S}}$ become equivalent in the slow-roll, large scale limit, where $\delta s$ is an isocurvature mode. We will present results for both comoving isocurvature perturbations in the following work.

Using the {\sc Pyflation} \cite{Huston:2009ac} code for computing inflationary perturbations we study the non-adiabatic pressure perturbation in two models of inflation containing two scalar fields. The numerical procedure and methods are described in detail in Ref.~\refcite{Huston:2011fr}.

First we will consider double-quadratic inflation, a model which is well-studied in the literature, with the potential
\be 
V(\varphi,\chi)=\frac{1}{2} m_{\varphi}^2\varphi^2+\frac{1}{2}m_\chi^2\chi^2\,.
\ee
We consider the case where $m_\chi=7 m_\varphi$ with $m_\varphi=1.395\times10^{-6}M_{\rm pl}$. The initial field values are $\varphi_0=\chi_0=12 M_{\rm pl}$, and these choices give $n_{\mathcal{R}}\simeq 0.937$.
\begin{figure}
\begin{center}\includegraphics[width=0.7\columnwidth]{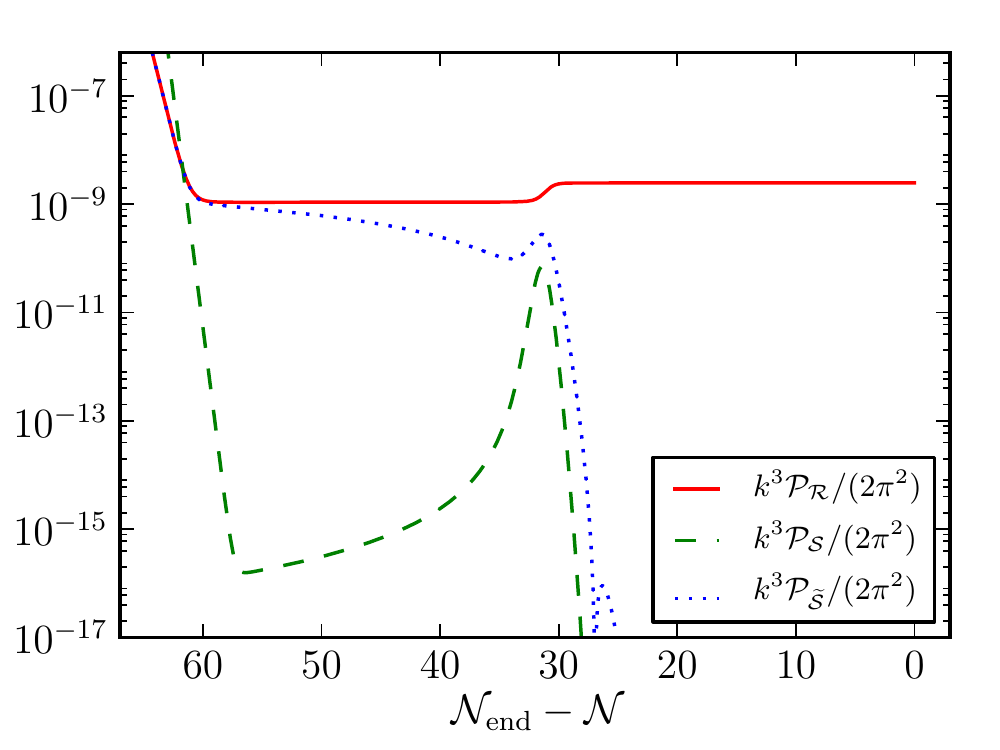}\end{center}
\caption{Double quadratic inflation: A comparison of the power spectra for ${\mathcal R}$ (red solid line), ${\mathcal S}$ (green dashed line) and $\widetilde{\mathcal S}$ (blue dotted line) at {\sc Wmap} pivot scale. The horizontal axis is the number of efolds (${\mathcal N}$) left until the end of inflation. Reproduced with permission from Huston and Christopherson (2012), their figure 3.}
\label{fig:hybquad_1}
\end{figure}

\begin{figure}
\begin{center}\includegraphics[width=0.7\columnwidth]{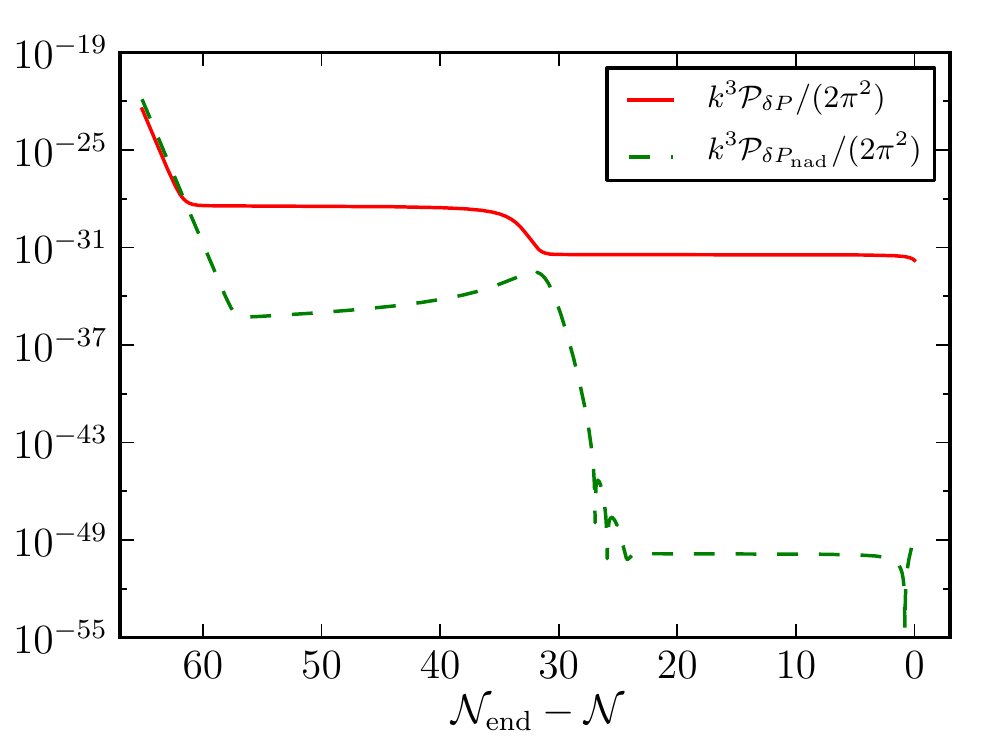}\end{center}
\caption{Double quadratic inflation: A comparison of the power spectra for the pressure (red solid line) and non-adiabatic pressure (green dashed line) perturbations. Reproduced with permission from Huston and Christopherson (2012), their figure 3.}
\label{fig:hybquad_2}
\end{figure}

In Fig.~\ref{fig:hybquad_1} we compare the power spectra of 
${\mathcal R}$, ${\mathcal S}$ and $\widetilde{\mathcal S}$ for this model. The dynamics of inflation are first governed by the $\chi$ field until around 30 efolds before the end of inflation when the $\varphi$ field takes over. We can see this from the curvature perturbation spectrum, where ${\mathcal R}$ is constant after the $k$-mode leaves the horizon, apart from around 30 efolds, where the switch in field domination causes a kick in the power spectrum. The $\widetilde{\mathcal S}$ spectrum drops off significantly after the change-over of field domination. The spectrum of the entropy perturbation, ${\mathcal S}$, quickly decreases as the mode leaves the horizon and grows again to reach its peak at the time of cross-over. It then decreases again rapidly, and at the end of inflation is orders of magnitude less than the curvature perturbation spectrum. 
The non-adiabatic pressure perturbation is plotted in Fig.~\ref{fig:hybquad_2}, and the evolution of $\delta P_{\rm nad}$ is somewhat similar to that of ${\mathcal S}$, as expected, with negligible non-adiabatic pressure perturbation at the end of inflation.

The second potential that we will consider is a hybrid inflation model with the potential, which has been studied in Refs.~\refcite{Avgoustidis:2011em,Kodama:2011vs}
\be 
V(\varphi,\chi)=\Lambda^4\Bigg[\Bigg(1-\frac{\chi^2}{v^2}\Bigg)^2+\frac{\varphi^2}{\mu^2}+\frac{2\varphi^2\chi^2}{\varphi_c^2 v^2}\Bigg]\,.
\ee
The parameter values are $v=0.1 M_{\rm pl}$, $\varphi_c=0.01 M_{\rm pl}$ and $\mu=10^3 M_{\rm pl}$. Then, $\Lambda$ can be normalized to {\sc Wmap} observations as $\Lambda=2.36\times10^{-4}M_{\rm pl}$, while the intial field calues are $\varphi_0=0.01M_{\rm pl}$ and $\chi_0=1.63\times10^{-9}$. These choices give $n_{\mathcal{R}}\simeq 0.932$. 

\begin{figure}
\begin{center}\includegraphics[width=0.7\columnwidth]{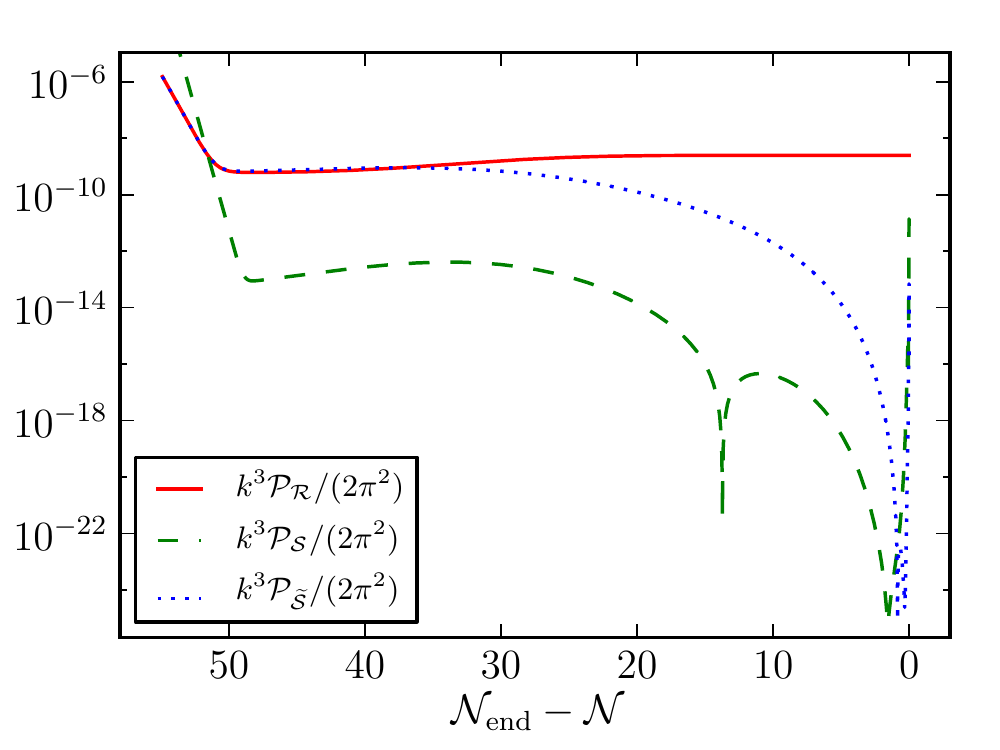}\end{center}
\caption{Double quartic inflation: A comparison of the power spectra for ${\mathcal R}$ (red solid line), ${\mathcal S}$ (green dashed line) and $\widetilde{\mathcal S}$ (blue dotted line) at {\sc Wmap} pivot scale. Reproduced with permission from Huston and Christopherson (2012), their figure 4.}
\label{fig:hybquart_1}
\end{figure}

The spectra for this model are plotted in Fig.~\ref{fig:hybquart_1}.

We can immediately see that the behavior of the entropy perturbations in this model is different to the previous case. ${\mathcal{R}}$ evolves slightly outside the horizon, while ${\mathcal{S}}$ slowly drops off as the mode exits the horizon before rising sharply towards the end of inflation. 
\begin{figure}
\begin{center}\includegraphics[width=0.7\columnwidth]{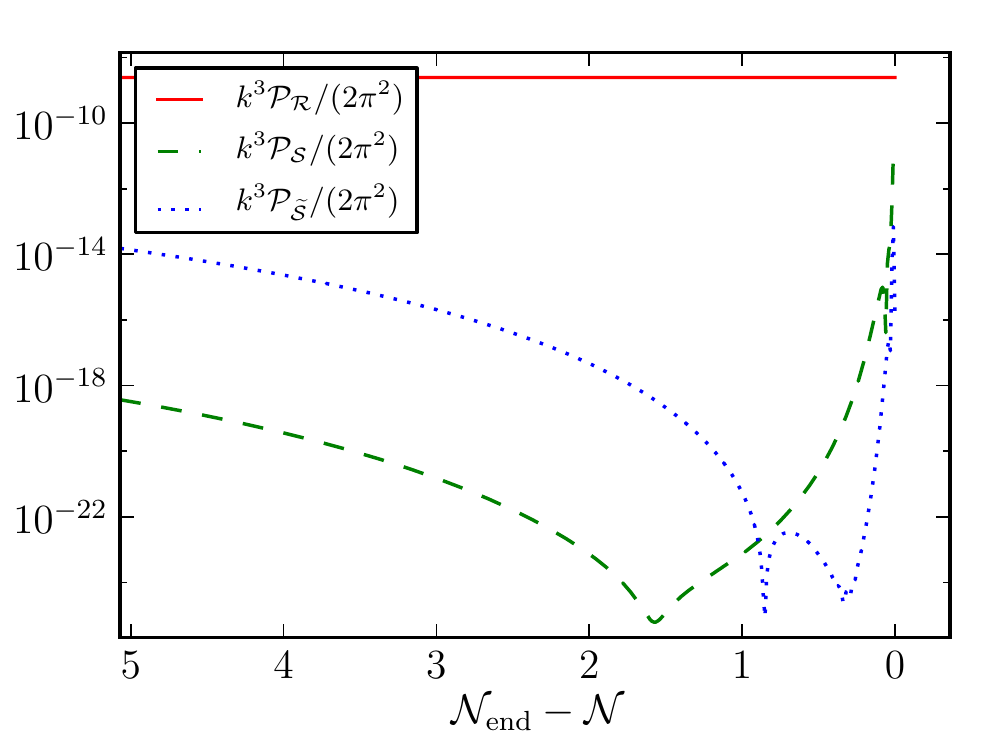}\end{center}
\caption{Double quartic inflation: A zoom of the last 5 efolds of Fig.~\ref{fig:hybquart_1}. Reproduced with permission from Huston and Christopherson (2012), their figure 6.}
\label{fig:hybquart_2}
\end{figure}
Zooming in to the last 5 efolds of evolution, as shown in Fig.~\ref{fig:hybquart_2}, we see that the entropy perturbations rebound towards the end of inflation. The amplitude of ${\mathcal S}$ at the end of inflation reaches a significant level, and is within a few percent of the curvature power spectrum.

\begin{figure}
\begin{center}\includegraphics[width=0.7\columnwidth]{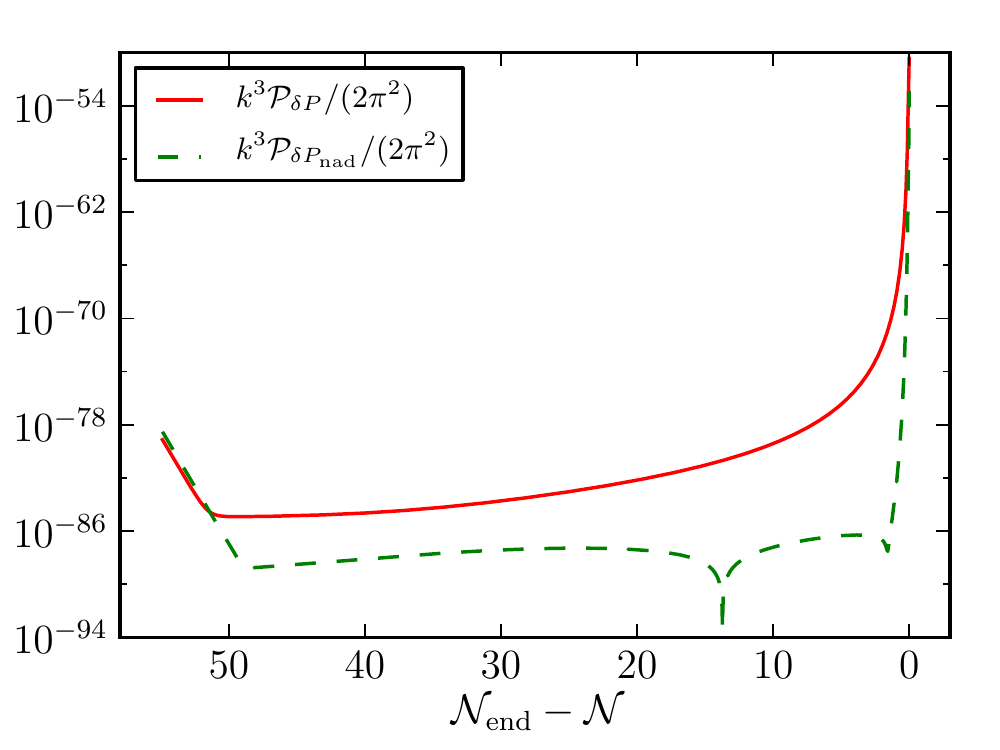}\end{center}
\caption{Double quartic inflation: A comparison of the power spectra for the pressure (red solid line) and non-adiabatic pressure (green dashed line) perturbations. Reproduced with permission from Huston and Christopherson (2012), their figure 7.}
\label{fig:hybquart_3}
\end{figure}

Similarly, the pressure perturbations evolve in a different way to those in the previous case of the double quadratic potential. As shown in Fig.~\ref{fig:hybquart_3}, we can see the pressure perturbation increasing throughout inflation, with the non-adiabatic pressure remaining almost constant until the last few efolds.

From the analysis of these two potentials, we can see that it is possible for two-field inflationary models to generate large non-adiabatic pressure perturbations at the end of inflation. While the entropy perturbations in the double quadratic potential are negligible, the double quartic model produces sizeable entropy perturbations.\cite{Huston:2011fr, Huston:2012dv} Of course, once inflation is over, the fields driving inflation must be converted into the standard model particles through a period of reheating. There is no agreed upon mechanism for reheating the universe. However, the significant amount of isocurvature present in the quartic model warrants a careful study of the dynamics after inflation. In fact, any prediction of statistics (such as the non-gaussianity of a given model) are open to change during the subsequent reheating phase.

The study of reheating is beyond the scope of this work -- preliminary results suggest that isocurvature perturbations do, in fact, decay during the reheating phase\cite{Huston:2013kgl}. However, this is very much an open question, and will be the subject of future work.

\section{Magnetic fields}
\label{sec:mag}

Magnetic fields are observed on all scales in the Universe, yet their primordial origin is still unexplained. Although astrophysical mechanisms could generate fields on small scales, their existence on the larger scales, and in voids, suggest a cosmological origin. One such mechanism for the generation of magnetic fields is by vorticity, which was first investigated by Harrison\cite{harrison}. Here, magnetic fields could be generated in a period around recombination by vorticity which naturally occurs at higher order in perturbation theory\cite{Gopal:2004ut, Takahashi:2005nd}. 

Thus, in the final section of this review article, I will focus on the magnetic field generation at second order in cosmological perturbation theory, presenting results derived in Ref.~\refcite{Nalson:2013jya}, using analytical techniques. There have been numerical studies addressing this problem: some work has focused on specific terms in the evolution equation\cite{Ichiki:2007hu, Maeda:2011uq}, while Ref.~\refcite{Fenu:2010kh} solved the full set of governing equations numerically. In Ref.~\refcite{Nalson:2013jya}, we undertook the first complete study using analytical techniques throughout. We derived the governing equations for the electric and magnetic field to second order in perturbation theory and then computed the power spectrum of the resulting magnetic field.

I will now briefly sketch the calculation of the second order magnetic field, however again stressing that one should refer to Ref.~\refcite{Nalson:2013jya} for full details. The governing equations for our system are the Einstein equations in a perturbed FLRW universe, as discussed in the previous sections of this article, along with the Maxwell equations for the electric and magnetic fields defined, respectively, as
\bea
{\mathcal{E}}^\mu &=& F^{\mu\nu} u_\nu\,,\\
{\mathcal{M}}^\mu &=& \frac{1}{2}\epsilon^{\mu\nu\lambda\delta}
u_\nu F_{\mu\delta}\,,
\eea
where $F^{\mu\nu}$ is the Faraday tensor. We expand the Maxwell equations order-by-order in perturbation theory, as described in Section~\ref{sec:modelling} for the gravitational equations, and further set the linear magnetic field and vorticity to zero, since they are not sourced at linear order. Additionally, we set the tensor perturbations, $h_{ij}$ to zero, and work only with scalar perturbations. Since our goal is to consider the generation of second order magnetic fields, the equation of interest is the evolution equation for ${\mathcal{M}}_2^i$
\bea 
 {{\MII}^{i}}'  +2 \H {\MII}^{i}   = {\epsilon}^{0 i j k} a^2
\Big[2 \Big({\partial}_{j}{\phi_1} -{{V_1}_{j}}'+ 2{V_1}_{j} \H\Big) {\EI}_{k} \nonumber\\
\qquad\qquad\qquad\qquad - {\partial}_{j}{{\EII}_{k}} 
 + 2 \mu_0 {V_1}_{j} a {\J_1}_{k}\Big]\,.
\eea
In order to close the system, the relevant matter-sector equations come from momentum conservation
\be
{{V_1}_{\alpha}}' +  (1 - 3 c_{\alpha}^2) \H {V_1}_{\alpha} + \phi_1 + \frac{1}{{\rho_0}_{\alpha} + {P_0}_{\alpha}}
\Big[{\dP_1}_{\alpha}  - \displaystyle\sum\limits_{\beta} {f_1}_{\alpha \beta }\Big] = 0\,,
\ee
where $c_\alpha^2$ is the adiabatic sound speed of the $\alpha$ fluid, i.e. $c_\alpha^2={P_0}_\alpha ' / {\rho_0}_\alpha '$ and $f_{\alpha\beta}$ is the momentum transfer between fluids \cite{Malik:2002jb}. This set of equations can then be solved, on specifying the matter content (e.g. baryons, photons) -- we omit this detail here.

Then, working in a radiation dominated regime, we can simplify the evolution equation for the magnetic field, using the governing equations, to become
\bea
 {{\MII}^{i}}'  + 2\H {\MII}^{i}   &=& 2a^2{\epsilon}^{0 i j k} 
\Bigg[\Big(\frac{\delta P_{1,j}}{c^2 \rho_0(1+w)} -(1-6c_{\rm s}^2+3w)\frac{V_{1,j}}{c}\Big){\EI}_{,k}\nonumber\\
&&\qquad\qquad\qquad\qquad-ac\mu_0{\J_1}_{,k}V_{1,j}-\frac{1}{2}{{\EII}_{k,j}}\Bigg]\,,
\eea
which we denote, in a shorthand, as 
\be 
{{\MII}^{i}}'  +2 \H {\MII}^{i}   = S^i\,.
\ee
This equation can be solved\cite{Nalson:2013jya} by transforming to Fourier space, expanding the magnetic field in a suitable basis and then considering the Fourier modes, to give the following simplified expression for the source term Fourier mode
 \bea
 \label{eq:Ssimp}
 S({\bf k},\eta)=\frac{a^2 k \bar{e}^j}{(1+w)\rho_0}
&&\int\frac{d^3{\bf k} \tilde{k}_j}{9\H^2(1+w)+2c^2\tilde{k}^2}\nonumber\\
&&\times 
 \Big[f(\tilde{k},\eta)\delta\rho_1(\tilde{\bf k},\eta)
 +g(\tilde{k},\eta)\delta P_{\rm nad 1}(\tilde{\bf k},\eta)\Big]
 \EI({\bf k}-\tilde{\bf k})\,,
 \eea
 with the following functions
 \bea
 f(\tilde{k},\eta)&\equiv& 2\H ac^2\mu_0(1+3w+6 c{\rm s}^2)J(\eta)\\
 &&\qquad-\Big(\H^2\Big[(1-6 c_{\rm s}^2+3w)(1+3w)-3c_{\rm s}^2(3c_{\rm s}^2+1)\Big]-c^2c_{\rm s}^2\tilde{k}^2\Big) E(\eta)\,,\nonumber\\
 g(\tilde{k},\eta)&\equiv& \frac{4}{c^2}\Big(3\H ac^2\mu_0 J(\eta)
 +\Big[3\H^2(3c_{\rm s}^2+1)+c^2\tilde{k}^2\Big]\Big)E(\eta)\,.
 \eea
We can see that the magnetic field spectrum therefore depends upon the energy density and the non-adiabatic pressure perturbations. The linear energy density solution is well-known, however there are two distinct sources for the non-adiabatic pressure perturbation, as we have seen in Section~\ref{sec:nonad}. These are, firstly, the non-adiabatic pressure perturbation arising from inflation driven by multiple scalar fields, which is imprinted onto the CMB as isocurvature perturbations, and secondly, the relative non-adiabatic pressure caused by the interaction between different fluids. We will consider these sources separately, and therefore obtain two contributions to the second order magnetic field power spectrum.

\begin{figure}
\begin{center}\includegraphics[width=0.8																																																																																																																											\columnwidth]{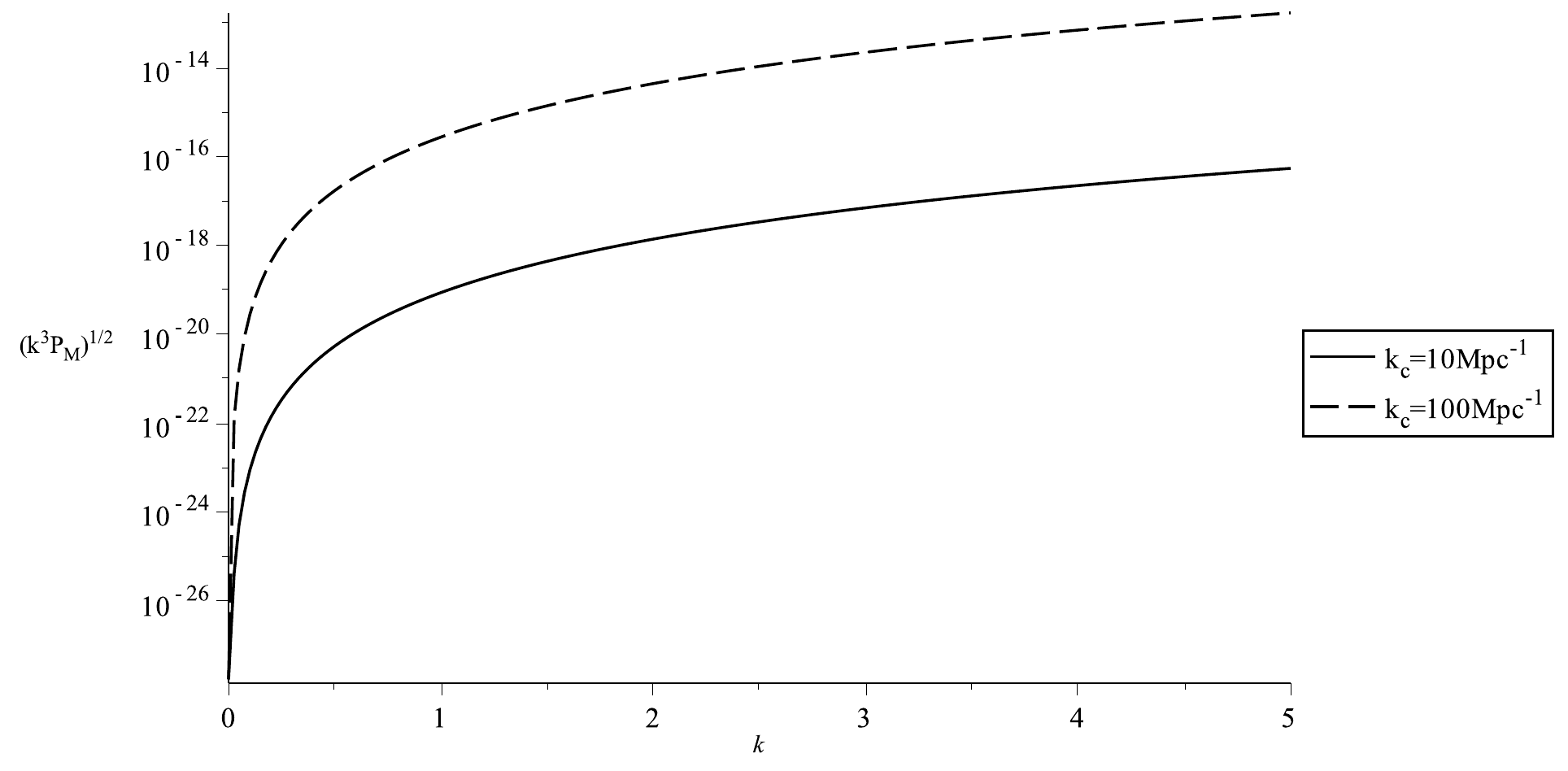}\end{center}
\caption{A plot showing the power spectrum of the magnetic field for two values of the small scale cutoff.}
\label{fig:mag}
\end{figure}

Omitting the lengthy calculation, we obtain the following expressions, where we have taken the leading order term and changed to S.I. units, allowing us to express the magnetic field in terms of Gauss: for the inflationary non-adiabatic pressure
\bea
 \sqrt{k^3 \Ps_{\M} (k, \eta)} &=& \frac{A E \eta_0}{32 \sqrt{2} (2\pi)^{3/2} \rho_0}  \left(\frac{k_c}{Mpc^{-1}}\right)^{\frac{13}{2}}  \left(\frac{\eta_c}{\eta}\right)^2 \nonumber\\
 &&\qquad\qquad\times\left[ \frac{32 }{135}  + \hat{\alpha} \frac{16}{27}\frac{k_c}{k_0} + \hat{\alpha}^2 \frac{8}{21} \left(\frac{k_c}{k_0}\right)^2 \right]^{\frac{1}{2}} \left( \frac{k}{k_c} \right)^4  \,,
 \eea
and for the relative non-adiabatic pressure, the magnetic field power spectrum is
\bea
 \sqrt{k^3 \Ps_{\M} (k, \eta)} &=& \frac{E A \eta_0}{32 \sqrt{2}(2\pi)^{3/2}\rho_0}  \left(\frac{k_c}{Mpc^{-1}}\right)^{\frac{13}{2}} \left(\frac{\eta_c}{\eta}\right)^2 \nonumber\\
 &&\qquad\qquad\times \left[\frac{32}{135} + \frac{32}{27}\frac{\hat{D}}{A}\left(\frac{k_c}{k_0}\right)^4 + \frac{24}{13}\frac{\hat{D}^2}{A^2} \left(\frac{k_c}{k_0}\right)^8\right]^{\frac{1}{2}}
\left(\frac{k}{k_c}\right)^4 \,.
\eea
Here, the constants are amplitudes of the energy density and non-adiabatic pressure power spectra, and of the linear electric field, and $\hat{\alpha}^2=\alpha(k_0)/\Big(1-\alpha(k_0)\Big)$, with $\alpha(k_0)$ defined in Eq.~(\ref{eq:alphadef}).
Of course, these expressions depend upon the small scale cutoff, $k_c$, as we saw for the vorticity calculation. Let's take the cutoff to be $k_c=10 Mpc^{-1}$, for which the inflationary contribution evaluated at $\eta=\eta_{\rm eq}$ becomes
\be 
\sqrt{ k^3 \Ps_{\M} } =  3.2\times 10^{-17}\Bigg[736.3\left(\frac{k}{10}\right)^8+515.4\left(\frac{k}{10}\right)^{10}-\frac{4}{315}\left(\frac{k}{10}\right)^{12}+\frac{4}{2835}\left(\frac{k}{10}\right)^{14}\Bigg]^{\frac{1}{2}}\,.
\ee
In order to estimate the size of the magnetic field, we will now evaluate the expressions above on cluster scales, of $k=1Mpc^{-1}$, today. 
For the inflationary non-adiabatic pressure contribution we obtain
\be
\sqrt{ k^3 \Ps_{\M} } \approx 5.9\times 10^{-27}G \,,
\ee
and for the relative non-adiabatic pressure
\be
\sqrt{ k^3 \Ps_{\M} } \approx 2\times 10^{-30}G \,.
\ee
Since the estimates are cutoff dependent, if we vary the cutoff slightly, we see that the results for the magnetic field strength vary a few orders of magnitude. In Fig.~\ref{fig:mag}, we show, for illustrative purposes, the field magnitude for two choices of the cutoff.

This result is the first analytical calculation of the second order magnetic field which takes into account all source terms in the evolution equation. Our result agrees with the numerical calculations presented in Ref.~\refcite{Fenu:2010kh}, thereby strengthening the numerical results. Additionally, the numerical calculations did not take into account any amplification of the signal by inflationary isocurvature, as we did in this work, since adiabatic initial conditions were used.

The magnetic field that we obtained is perhaps too small to be the source of the primordial seed field, however there is still some work to be done in order to definitively rule this method out as the primordial seed source. Since the power spectrum is rising on small scales, it is possible that some power could move coherently from short to large scales. A complete calculation including the small scale effects could lead to an enhanced result. Additionally, in this work we neglected tensor perturbations, and linear magnetic fields. In particular, it would be interesting to study how the estimate for the primordial seed field would change if we were to include a linear magnetic field; these non-linear effects might change predictions from, for example, inflationary magnetogenesis mechanisms. These effects will all be studied in a future publication.

\section{Discussion}
\label{sec:discuss}

Cosmological perturbation theory is an important technique in theoretical cosmology allowing us to make and test predictions of our theory of cosmology. The linear theory well describes observations of inhomogeneities and anisotropies of the CMB and large scale structure. Recently, predictions made using higher order theory have been tested, such as the non-Gaussian signature of the primordial perturbations.
In this review article, I have summarized various complementary aspects of cosmological perturbation theory, all with the underlying theme of the presence of non-adiabatic pressure perturbations.

First, I introduced Newtonian perturbation theory and presented the the Euler, energy conservation and Poisson equations.  These can then be combined into a single, second order differential equation governing the system, from which the physics becomes more apparent. Then, I showed how to extend this to fully relativistic perturbation theory around FLRW, both at linear and second order. This allows us to model fluids moving at an appreciable fraction of the speed of light, or scales that are comparable to the horizon size, which are both important for cosmology. The main difference between the Newtonian and relativistic theories is the so-called {\emph{gauge problem}}. Since the process of splitting the spacetime into background and perturbations is not a covariant process, there exist spurious gauge modes. These are not physical, but are extra degrees of freedom, which can be removed by constructing gauge invariant variables. Equivalently, one can make a choice of gauge, which also removes the gauge freedom. I then presented the governing equations in the uniform curvature gauge.

After this introduction, I considered vorticity, both in Newtonian fluid mechanics, and in relativistic cosmology. As has been known for some time, vorticity can be sourced by entropy gradients in classical fluid mechanics. I extended this in some recent work to a cosmological setting, showing that in cosmological perturbation theory the vorticity is sourced at second order by a coupling between linear energy density and entropy perturbations. 

Then, I presented some work on non-adiabatic pressure, or entropy, perturbations focusing on two different physical systems. The first of these was the relative entropy perturbation in the cosmic fluid, which exists between the relativistic and non-relativistic species. This has a non-vanishing signature, even starting with adiabatic initial conditions, which peaks around matter radiation equality. The second concerned isocurvature perturbations in two-field inflationary models. We found that, of the two models studied, sizeable entropy perturbations were only generated in the double quartic model. It should be noted that these predictions from inflation may change when a reheating mechanism is studied, however this is beyond the scope of the current work.

Finally, I reviewed some recent work on magnetic field generation. At second order in perturbation theory, magnetic fields are generated, again from density and entropy perturbations. Through this fully analytical work, we find a magnetic seed field that has the scale dependence $k^4$. We find agreement with recent numerical work, however our approach also allow for non-adiabatic initial conditions. The magnetic field we generate is, unfortunately, too small to provide the seed field that is amplified by, e.g., battery mechanisms to become the fields observed on cluster scales today. However, there is much work still to be done, in particular in performing the small scale calculation, or considering the possible amplification of linear fields by second order effects.

I have shown, and argued, how these different aspects of cosmological perturbation theory, both at first and second order, could have important observational consequences. In particular, we have seen the importance of non-adiabatic pressure, or entropy, perturbations. Not only do they have observational consequences in their own right, but they can generate vorticity and magnetic fields at higher order.
This provides strong motivation for future work investigating the primordial perturbations, and for trying to understand the physics of inflation and the early universe.

\section*{Acknowledgements}

The author is extremely grateful to Iain Brown, Juan Carlos Hidalgo, Ian Huston, Karim Malik, David Matravers and Ellie Nalson for the enjoyable collaborations on which this work is based. AJC is funded by the Sir Norman Lockyer Fellowship of the Royal Astronomical Society, and thankfully acknowledges the hospitality of several institutions in the USA when this work was presented, as well as the Astronomy Unit, QMUL for ongoing collaboration and hospitality.

\bibliographystyle{ws-ijmpd.bst}
\bibliography{christopherson_ijmpd.bbl}

\end{document}